\documentclass[copyright,creativecommons]{eptcs}
\usepackage[utf8]{inputenc}

\providecommand{\event}{QPL 2020} %
\usepackage{underscore}           %

\usepackage[T1]{fontenc}
\usepackage{mathpartir}
\usepackage{amssymb}
\usepackage{amsmath}
\usepackage{braket}
\usepackage{color}
\usepackage{xspace}
\usepackage{cleveref}
\crefformat{section}{\S#2#1#3}

\usepackage{etoolbox}
\apptocmd{\sloppy}{\hbadness 3000\relax}{}{}

\hypersetup{
	pdfauthor={Kartik Singhal},%
	pdftitle={Quantum Hoare Type Theory},%
	pdfsubject={QPL 2020}
}

\definecolor{dkgreen}{rgb}{0,0.35,0}

\usepackage{listings}
\lstdefinelanguage{QHaskell}{
	language     = Haskell,
	morekeywords = {Qbit, U, C, Pure},
}

\newcommand{\type}[1]{\textrm{\textbf{#1}}}
\newcommand{\kw}[1]{\,\mathrm{\texttt{#1}}\,}

\definecolor{dRed}{rgb}{0.85, 0.0, 0.0}
\definecolor{dBlue}{rgb}{0.0, 0.0, 0.85}

\newcommand{\chkcolor}{dBlue}
\newcommand{\syncolor}{dRed}
\newcommand{\chk}{\,\textcolor{\chkcolor}{\Leftarrow}\,}
\newcommand{\uncoloredsyn}{{\Rightarrow}}
\newcommand{\syn}{\,\textcolor{\syncolor}{\uncoloredsyn}\,}

\providecommand{\thisvolume}[1]{this volume of EPTCS, Open Publishing Association}

\newcommand{\mailtodomain}[1]{\href{mailto:#1@cs.uchicago.edu}{\nolinkurl{#1}}}

\title{Quantum Hoare Type Theory: Extended Abstract}
\author{
Kartik Singhal \qquad\qquad John Reppy
\institute{University of Chicago}
\email{\{\mailtodomain{ks}, \mailtodomain{jhr}\}@cs.uchicago.edu}
}

\def\authorrunning{Kartik Singhal \& John Reppy}
\def\titlerunning{Quantum Hoare Type Theory}

\begin{document}

\maketitle

\begin{abstract}
As quantum computers become real, it is high time we come up with effective techniques that help programmers write correct quantum programs.
In classical computing, formal verification and sound static type systems prevent several classes of bugs from being introduced. There is a need for similar techniques in the quantum regime.
Inspired by Hoare Type Theory~\cite{htt-jfp} in the classical paradigm, we propose Quantum Hoare Types by extending the Quantum IO Monad~\cite{qio} by indexing it with pre- and postconditions that serve as program specifications.
In this paper, we introduce Quantum Hoare Type Theory (QHTT), present its syntax and typing rules, and demonstrate its effectiveness with the help of examples.

QHTT has the potential to be a unified system for programming, specifying, and reasoning about quantum programs.
This is a work in progress.\footnote{An updated report is available~\cite{singhal2020quantum} at the time of this publication.}
\end{abstract}

\section{Introduction}
It is difficult to reason about the correctness of quantum programs. Sound static type systems help prevent a huge class of bugs from occurring but since the realm of quantum programming is still new there is not a lot of consensus on what kind of types make the most sense. Further, it is unclear how much they help programmers reason about the semantic properties associated with the quantum algorithms that they are implementing.

Recent work~\cite{huang2018,huang2019} as part of EPiQC\footnote{EPiQC: Enabling Practical-Scale Quantum Computation:  \url{https://epiqc.cs.uchicago.edu}} has identified several classes of bugs in quantum programs and proposed approaches to tackle them. The technique that holds the most promise is assertion checking using preconditions and postconditions. But assertions are usually checked dynamically during runtime, which can be wasteful of precious quantum computing or simulation resources.

As programming languages researchers, we think it will be better to encode such assertions into a static type system both for formal verification and to aid the programmers in writing correct programs from the start. Inspired by the use of Hoare triples in the verification of imperative programs and building on the idea of Hoare Type Theory (HTT) for classical programming languages~\cite{htt-jfp}, we extend the Quantum IO Monad interface~\cite{qio} and propose Quantum Hoare Type Theory aimed at enabling both sound static type checking and formal verification of quantum programs.

The main idea is that instead of just indexing the QIO monad type with the type of computation result, we can also index it with preconditions and postconditions so as to integrate Hoare-style reasoning into the type system itself. The resulting Quantum Hoare Type, $\{P\} x{ : }A \{Q\}$, specifies the preconditions, $P$, that hold on the quantum state before execution; the result, $x$, and its type, $A$; and, the postconditions, $Q$, that are true for the quantum state after successful execution. In this way the effectful quantum fragment of the program is effectively encapsulated inside the Quantum Hoare monad. Our theory, like HTT, allows the usual equational reasoning for the pure classical fragments of the program and uses syntax-directed type checking for generating strongest postconditions for the quantum fragment.

This paper is organized as follows. We discuss some related work and background in the next section, \cref{sec:bg}; specifically, Hoare Type Theory in the classical setting (\cref{sec:htt}) and then the Quantum IO Monad interface (\cref{sec:qio}). Then we introduce our contributions in merging these ideas together for specification and enforcement of useful semantic properties in section \cref{sec:qhtt} along with its syntax (\cref{sec:syntax}), typing rules (\cref{sec:typing}) and, some examples (\cref{sec:examples}). Finally, we conclude and share ideas for future work in the final section (\cref{sec:conclusion}).

\section{Related Work and Background}
\label{sec:bg}
Previous work, such as Proto-Quipper~\cite{ross2015,rios2017} and QWire~\cite{Paykin2017,rand2017,rand2018}, utilize a linear type system and dependent types to enforce a small subset of semantic properties, such as the no-cloning restriction and whether a unitary gate is of the right dimension. These advances in quantum type systems, although helpful, still fall short in encoding and enforcing even more useful properties that one would like to be able to express for the purpose of verification.

Our approach builds upon previous work in reasoning about quantum programs such as Quantum Weakest Preconditions~\cite{dhondt2006} and Quantum Hoare Logic~\cite{ying2012} in the spirit of Hoare~\cite{hoare1969} and Dijkstra~\cite{dijkstra1976} but attempts to bring those reasoning techniques into the type system. The hope is that programmers will be able to encode some of the semantic properties that they expect of their programs as specifications in their code and type checking will ensure correctness of some of those properties. In the classical setting, Hoare Type Theory~\cite{htt-jfp} accomplishes exactly this goal. Our attempt is to merge these ideas for the quantum realm.

In the rest of this section, we provide background on the core ideas from existing literature that form the foundation for our work --- Hoare Type Theory (HTT) and the Quantum IO Monad interface. We assume background in basics of quantum computing and refer the reader to Nielsen's excellent series of essays~\cite{matuschak2019} for a first introduction and to the standard textbook~\cite{nielsen2010} for advanced material.

\subsection{Hoare Type Theory}
\label{sec:htt}
A Hoare type, $\Delta.\Psi.\{P\}\; \texttt{x}:A\; \{Q\}$, encodes preconditions and postconditions in the same spirit as Hoare triples to allow both specification and verification of effectful classical programs. It can be read as `for a stateful computation executed in a heap that satisfies precondition $P$, return a value of type $A$ in a heap that satisfies postcondition $Q$.' The contexts, $\Delta$ and $\Psi$, contain the variables and heap variables that may appear in both $P$ and $Q$. In this presentation, we will omit the contexts in the Hoare type when they are unneeded. We show an example to demonstrate the expressiveness of a Hoare type --- the \texttt{alloc} primitive from HTT~\cite{htt-jfp}:
\begin{mathpar}
	\texttt{alloc} : \forall \alpha. \Pi x:\alpha. \{\texttt{emp}\}\; y:\texttt{nat}\; \{y \mapsto_{\alpha} x\}
\end{mathpar}

This type specifies that \texttt{alloc} is a polymorphic function that takes as input a variable, $x$, of any simple type, $\alpha$, that is executed in an empty heap (meaning it does not affect existing heap), returns a new location bound by a fresh variable, $y$, of type \texttt{nat} and initializes it with the supplied value $x$ of type $\alpha$.

Type checking in HTT involves generation of strongest postconditions at each step of the program. Verification is a two-step process: the first phase does basic type checking and verification condition generation which is decidable, the second phase needs to show the validity of the generated verification conditions, which can be undecidable. This second phase can be deferred to an automated theorem prover.

We provide more details when we introduce QHTT in section~\cref{sec:qhtt}. An accessible introduction to Hoare Type Theory is available in the lecture notes by Perconti~\cite{perconti2012}.

\subsection{Quantum IO Monad}
\label{sec:qio}

The Quantum IO (QIO) monad~\cite{qio} is a purely functional interface for quantum programming that provides a separation between the unitary (reversible) and non-unitary (irreversible) fragments of quantum computation. It provides isolation of quantum effects inside a monad similar to what we need to provide in our Quantum Hoare monad.

QIO interface was developed as a library for the Haskell programming language but its design was influenced~\cite{asg2010} by the category \textbf{FQC} of finite quantum computations (where computations are interpreted as superoperators) as explored by the authors in previous work~\cite{fqc06}. This is the most we will say about the relation of QIO monad to category theory in this paper.

Here we discuss the relevant bits from the QIO interface. QIO provides a type for referring to qubits, \texttt{Qbit}; a type for unitary operations, \texttt{U}; and, the \texttt{QIO} type operator which is a monad indexed by the result of the quantum computation. The primitive quantum operations defined using the QIO monad below (Haskell syntax) are:
\begin{lstlisting}[language=QHaskell]
mkQbit :: Bool -> QIO Qbit       -- initialization
applyU :: U -> QIO ()            -- apply a unitary to quantum state
measQbit :: Qbit -> QIO Bool     -- measurement
\end{lstlisting}

Further, \texttt{U} is defined as a monoid with neutral element, \texttt{mempty}, and operation, \texttt{mappend}, that encodes reversible operations on quantum state. The core unitary primitives that we will need are:

\begin{lstlisting}[language=QHaskell]
rot :: Qbit -> ((Bool, Bool) -> C) -> U
cond :: Qbit -> (Bool -> U) -> U
\end{lstlisting}

\texttt{rot} takes a qubit and a two-by-two complex-valued unitary matrix (represented as a function from the matrix indices to $\mathbb{C}$) and lets one define arbitrary rotation on a single qubit. \texttt{rot} can then be used to define standard gates such as the Hadamard, H, and the Pauli gates, X, Y and Z. \texttt{cond} is the unitary conditional that is used to perform branching unitary operations. For the present paper, we use a simpler conditional, \texttt{ifQ}, that is defined as follows:

\begin{lstlisting}[language=QHaskell]
ifQ :: Qbit -> U -> U
ifQ q u = cond q (\x -> if x then u else mempty)
\end{lstlisting}

That is, \texttt{ifQ} acts as the standard control operator in the quantum circuit model that runs its second argument based on the truth value of its first argument. Then, it is easy to define controlled gates such as CNOT with the expression $\kw{ifQ q1 (X q2)}$.

Quantum state in the QIO interface is modeled as a normalized vector that stores pairs representing complex amplitudes associated with each basis state (represented as a map from Qbit to Bool types). The operations defined over the vector class ensure that the quantum state is kept normalized throughout.

QIO further ensures that the monoidal structure of unitary operations lets one run a unitary operation over the complete quantum state. We elide details here and refer the reader to \cite{qio} for more.

\section{Quantum Hoare Type Theory}
\label{sec:qhtt}

In this section, we introduce the Quantum Hoare Type inspired by the QIO monad and its type theory (QHTT) by replacing the classical effectful portion of Hoare Type Theory with quantum effects. The core idea is to encapsulate any quantum effect inside a monadic Quantum Hoare type, $\Delta. \Psi .\{P\} x{ : }A \{Q\}$, and formalize reasoning of quantum effects using strongest postconditions in a similar fashion as HTT in the classical setting.

The next two sections discuss the syntax and typing of QHTT. The last section show examples written in QHTT.

\subsection{Syntax}
\label{sec:syntax}

\Cref{fig:syntax} shows the syntax of Quantum Hoare Type Theory. Our presentation closely follows that of HTT~\cite{htt-jfp}.

\begin{figure}[t]
	\begin{tabular}{lrcl}
		\textit{Types} & $A, B, C$ & ::= &
		\begin{minipage}[t]{0.6\columnwidth}%
			$ \type{1} \mid \type{Bool} \mid \type{Qbit} \mid \type{U} \mid \type{Pure} \mid A \otimes B \mid \Pi x{ : }A.B \mid \Delta. \Psi .\{P\} x{ : }A \{Q\}$
		\end{minipage}\\ \\
		\textit{Assertions} & $P, Q, R$ & ::= &
		\begin{minipage}[t]{0.5\columnwidth}%
			$ \top \mid \bot \mid P \wedge Q \mid P \vee Q \mid P \supset Q \mid \neg P \mid \exists x{:}A.P \mid \forall x{:}A.P \mid \exists h{:}\kw{heap}.P \mid \forall h{:}\kw{heap}.P \mid \kw{Id}_A(M, N) \mid \kw{HId}(H, G) \mid \kw{indom}(H,M) $
		\end{minipage}\\ \\
		\textit{Q. Heaps} & $H, G$ & ::= & $ h \mid \kw{empty} \mid \kw{upd}(H, M, N) $\\ \\
		\textit{Elim terms} & $K, L$ & ::= & $ x \mid K M \mid M : A $\\
		\textit{Intro terms} & $M, N, O$ & ::= & $ K \mid () \mid \lambda x.M \mid \kw{do} E \mid \kw{true} \mid \kw{false} $\\ \\
		\textit{Q. Commands} & $c$ & ::= &
		\begin{minipage}[t]{0.6\columnwidth}%
		    $ \kw{mkQbit} M \mid \kw{measQbit} M \mid \kw{applyU} M \mid \kw{if} M \kw{then} N \kw{else} O$
		\end{minipage}\\
		\textit{Computations} & $E, F$ & ::= & $ \kw{return} M \mid x \leftarrow K; E \mid x \Leftarrow c; E \mid x =_A M; E $\\ \\
        \textit{Variable context} & $\Delta$ & ::= & $ \cdot \mid x : A, \Delta $ \\
		\textit{Q. Heap context} & $\Psi$ & ::= & $ \cdot \mid h, \Psi$ \\
		\textit{Assertion context} & $\Gamma$ & ::= & $ \cdot \mid P, \Gamma$ \\
	\end{tabular}
	\caption{Syntax of Quantum Hoare Type Theory}
	\label{fig:syntax}
\end{figure}

\paragraph{Types} We include primitive types for unit, booleans, qubits and unitary operations, and a type \type{Pure} for representing pure quantum state vectors (complex valued vectors in Hilbert space); these correspond to similar types in the QIO work. There are also type constructors for pairs, dependent functions and the Hoare type from HTT.

\paragraph{Assertions}
Apart from the usual first order logic, we have assertions for reasoning about propositional equality of terms, $\kw{Id}_A(M, N)$, and heaps, $\kw{HId}(H, G)$.

As done in HTT, some common convenience assertions can be defined using the base primitives. Particularly, $\kw{emp}$ denotes that the current heap, say $h$, is empty, that is, $\kw{HId}(h, \kw{empty})$. $M \mapsto N$ represents the only mapping in the (singleton) heap. $M \hookrightarrow N$ says that looking up $M$ in the heap returns $N$ which can be more explicitly written as $\kw{seleq}(H, M, N)$.

\paragraph{Quantum State}

We use the terms quantum state and quantum heaps interchangeably throughout. A quantum heap is a partial function from qubits (that can in turn be thought of as locations, represented using natural numbers) to at most one quantum state vector.

For the purpose of the type theory, heaps are functional, so that, for example, $\kw{upd}(H, M, N)$ returns a new heap after updating heap $H$ at location $M$ with $N$.

\paragraph{Terms}
The terms are divided into introduction and elimination sorts for the purpose of bidirectional typechecking like in HTT.

The pure fragment allows higher-order functions while the impure (effectful) fragment is encapsulated within the monadic \texttt{do} $E$ construct and supports a simple imperative quantum language inspired by QIO constructs. It is worth noting that \texttt{do} $E$ represents suspended computation and is considered pure.

\paragraph{Quantum Commands}
These are the quantum-specific effectful commands from QIO for initialization, unitary application and measurement of qubits. We also support classical control using the \texttt{if} $M$ \texttt{then} $N$ \texttt{else} $O$ construct similar to HTT.

\paragraph{Computations} The computations are the usual monadic return and three sequencing operations that bind their result to a variable similar to HTT. The first, $x \leftarrow K$, executes a suspended computation (such as applying a function that includes a \texttt{do} $E$ in its body); the second, $x \Leftarrow c$ executes a primitive command $c$; and the last, $x =_A M$, is just syntactic sugar for let-binding.

\paragraph{Contexts} Contexts are modelled exactly the same way as in HTT. The only difference is that instead of classical heaps, we have quantum heaps.

\subsection{Typing Rules}
\label{sec:typing}

In this section, we provide some details about type checking in HTT and how it is different in the quantum case.

\subsubsection{Judgments}
We explain the main judgments used in HTT that are required to understand the typing rules in the next section. The type system is designed to be bidirectional and syntax directed and hence includes separate judgments for intro and elim terms. We are able to use them unchanged for QHTT. A recent survey provides accessible introduction to bidirectional typing~\cite{dunfield2019bidirectional}.

The type checking process also involves generating canonical forms for each term. A canonical form in HTT means a beta-normal (containing no beta-redexes) and eta-long (all intro terms are eta-expanded) form. We elide much detail here.

As usual, elim terms synthesize types ($\syn$) and intro terms are checked against the given type ($\chk$). Canonical forms are always synthesized. For the pure fragment:

\begin{center}
\begin{tabular}{ll}
    $\Delta \vdash K \syn A [N']$ & Elim term K has type A and canonical form $N'$ \\
    $\Delta \vdash M \chk A [M']$ & Intro term M has type A and canonical form $M'$
\end{tabular}
\end{center}

Note that the blue colored $\chk$ symbol we use in the typing rules lives in a different syntactic category from the $\Leftarrow$ symbol used in the syntax (\cref{fig:syntax}) that binds the output of primitive commands.

The judgments for computations involve synthesizing the strongest postconditions and checking whether a given postcondition applies:

\begin{center}
\begin{tabular}{ll}
    $\Delta; P \vdash E \syn x:A.Q[E']$ & Computation E with precondition P has \textit{strongest} postcondition Q\\ & and returns value x of type A. Its canonical form is $E'$. \\
    $\Delta; P \vdash E \chk x:A.Q[E']$ & Computation E with precondition P has postcondition Q\\ & and returns value x of type A. Its canonical form is $E'$.
\end{tabular}
\end{center}

Even though, we have shown the full forms of these judgments above, we will omit the canonical forms while presenting typing rules in the next section as they crowd the rules and do not affect the insight to be gained from them.

Finally, the judgment $\Delta; \Psi; \Gamma_1 \Longrightarrow \Gamma_2$ encodes the sequent calculus for the assertion logic. Recall that $\Delta$, $\Psi$, and $\Gamma$ denote the variable context, the quantum heap context and the assertion context, respectively.

The type system includes rules for primitive effectful commands and those for structuring composition such as monadic unit and bind. We reuse most of the HTT rules except those for primitives for quantum effects that need to specify the strongest postconditions for each primitive, we discuss that next.

\subsubsection{Strongest Postconditions}
Here we show the strongest postconditions for the primitive quantum commands of QHTT encoded in their typing rules. Given a quantum command for initialization, unitary application or measurement, its strongest postcondition is an assertion that most precisely captures the relationship between the initial state and the modified state after the execution. These rules along with the \texttt{consq} (consequent) rule of HTT (that ensures weakening of the given strongest postcondition to an arbitrary postcondition) work together to ensure composition of verification.

We need to use the relational composition connective from HTT ($P \circ Q$) that captures how heap evolves with computation. It basically reads: $Q$ holds of the current heap which is obtained after modification of a prior heap for which P holds.

We also use HTT's difference operator ($\multimap$) below that captures changes to only the interesting fragment of the heap without modifying the rest of the heap.

We can now state some typing rules:

\paragraph{Initialization}
$x \Leftarrow \kw{mkQbit} M; E$

\begin{mathpar}
  \inferrule
  {\Delta \vdash M \chk \type{Bool}
  \\ \Delta, x : \type{Qbit}; P \circ (x \mapsto \kw{state}(M)) \vdash E \syn y:B . Q}
  {\Delta; P \vdash x \Leftarrow \kw{mkQbit} M; E \syn y:B. (\exists x:\type{Qbit}.Q)}
\end{mathpar}
where x is fresh and \texttt{state} is a function that translates a classical representation of quantum state (such as boolean here) to the equivalent quantum state vector.

Initialization can only be performed for a term $M$ that can be type checked as a \type{Bool}. Then, we look at the rest of the computation, $E$, in a context that adds $x$ of type \type{Qbit} under a precondition that extends the previous precondition, $P$, with the strongest postcondition for initialization, that is, the newly bound variable $x$ points to the representation of a new qubit state; this should synthesize the strongest postcondition (with respect to the expanded precondition in the context), $Q$, for $E$.

To avoid dangling variables, we need to existentially quantify $x$ in the postcondition, $Q$, of the conclusion.

\paragraph{Measurement}
$x \Leftarrow \kw{measQbit} M; E$

\begin{mathpar}
  \inferrule
  {\Delta \vdash M \chk \type{Qbit}
  \\ \Delta; \Psi; P \Longrightarrow (M \hookrightarrow -)
  \\ \Delta, x:\type{Bool}; P \circ ((M \mapsto -) \multimap \kw{emp}) \vdash E \syn y:B . Q}
  {\Delta; P \vdash x \Leftarrow \kw{measQbit} M; E \syn y:B. (\exists x:\type{Bool}.Q)}
\end{mathpar}

The measurement rule can be read in a similar way as the previous rule. But in measurement, we need to additionally prove that the verification condition (written as a sequent), the precondition $P$ implies that the qubit $M$ is allocated, holds in the heap context $\Psi$. The strongest postcondition here is that the fragment of heap that only refers to location M becomes empty.

As may be apparent, initialization is analogous to \texttt{alloc} and measurement to \texttt{dealloc} and lookup primitives of HTT. We are working out precise details for unitary application, but at a high-level, they involve ensuring that the given unitary term actually represents a unitary matrix and updating the quantum state with the result of unitary application. This involves incorporating the monoidal structure of unitaries into our theory (as is done in QIO). This step also involves generating verification conditions that cannot be checked during this first phase of typechecking. We are also considering more tractable alternatives such as the sum-over-paths action semantics~\cite{amy18} based on the Feynman path integral formulation.\footnote{We ended up taking a different approach in subsequent work~\cite{singhal2020quantum}.}

\subsection{Sample programs}
\label{sec:examples}

In this section, we show some simple examples from the QIO work~\cite{qio} translated in our language and annotated with their program specifications using their Quantum Hoare types. An observation is that translating the programs was a very simple process except for coming up with the right specifications for the programs. We provide some commentary on assertions specified for these programs.

At the end of this section, we show a sample verification using QHTT for Bell pair generation.

\subsubsection*{Hello Quantum World}
\begin{lstlisting}[language=QHaskell]
hqw : {emp} r : Bool {emp /\ Id(r, false)}
    = do q <= mkQbit false;
         measQbit q
\end{lstlisting}

In this trivial program where we initialize a new qubit with \texttt{false} and then immediately measure it, we assert that the result is equal to \texttt{false}. The complete specification also implicitly says that the existing quantum state is not affected in any way as both the pre- and postcondition include the \texttt{emp} assertion.

\subsubsection*{Coin Toss}

\begin{lstlisting}[language=QHaskell]
rnd : {emp} r : Bool {emp}
    = do q <= mkQbit false;
         applyU (H q);
         measQbit q
\end{lstlisting}

In the case of quantum coin toss, the result can be in either of its two boolean values \texttt{false} or \texttt{true}, hence, we need not specify any special postcondition as type checking the result will be sufficient for correctness.

\subsubsection*{Testing Bell Pair}
\begin{lstlisting}[language=QHaskell]
testBell : {emp} (a, b) : (Bool, Bool) {emp /\ Id(a, b)}  -- a,b \in {true, false}
         = do qa <= mkQbit false;
              applyU (H qa);
              qb <= mkQbit false;
              applyU (ifQ qa (X qb));
              (measQbit qa, measQbit qb)
\end{lstlisting}

Here, we are asserting that the returned booleans $a$ and $b$ hold the same value, which is what we expect from the first Bell state prepared in this program.

\paragraph{Verification}
We would like to describe how QHTT lets us verify the correctness of the above program based on the specification provided as the type of the function. For illustration, we show the assertions annotated as comments as we step through the \texttt{testBell} program below. We have elided specific details about unitary application as they are still being worked out.

\begin{lstlisting}[language=QHaskell]
testBell : {emp} (a, b) : (Bool, Bool) {emp /\ Id(a, b)}
              -- P0: emp
         = do qa <= mkQbit false;
              -- P1: P0 \o (qa |-> |0\>)
              applyU (H qa);
              -- P2: P1 \o ((qa |-> |0\>) -o (qa |-> |+\>))
              qb <= mkQbit false;
              -- P3: P2 \o (qb |-> |0\>)
              applyU (ifQ qa (X qb));
              -- P4: P3 \o ((qa |-> |+\>, qb |-> |0\>) -o (qa, qb) |-> |\Phi+\>)
              (measQbit qa, measQbit qb)
              -- P5: P4 \o ((qa |-> -) -o emp) \o ((qb |-> -) -o emp)
\end{lstlisting}

In the assertion P2, we show that only the portion of the quantum state referred to by the \texttt{applyU} command is affected. Similarly, in P4 we see that only the relevant portion of the heap containing the control and target qubits used in the CNOT operation are affected. In this case, we also needed to combine the two qubit state as they are now entangled. The last assertion, P5, merely encodes the strongest postcondition for measurement for each of the two qubits.

Finally, the \texttt{consq} rule of HTT ensures that the assertion P5 leads to the specified postcondition in the type of the program.

We show some two more examples in the appendix for a modular version of Bell pair generation (\Cref{app:mbp}) that shows use of higher-order features of QHTT and another for teleportation protocol (\Cref{app:teleport}). Both of these involve specifying assertions over quantum state whose theory we still need to work out.

\section{Conclusions and Future Work}
\label{sec:conclusion}
In this paper, we have described our ongoing work on Quantum Hoare Type Theory, which modifies HTT for quantum computing using ideas from the Quantum IO monad work. This approach has the potential to be a unified system for programming, specification, and reasoning about quantum programs. This is active work in progress and more details will be available in a forthcoming technical report~\cite{singhal2020quantum}.

\bigskip

There are several avenues of future work to explore:

\paragraph{Mechanization} There are multiple implementations of HTT in Coq such as Ynot~\cite{ynot2008}. We are similarly working on mechanizing QHTT in Coq for higher assurance of the usefulness and soundness of our theory. This will also enable us to extract verified circuits in a lower level quantum language such as OpenQASM for execution on real machines.

\paragraph{Quantum Assertion Logic} An obvious need is to come up with an assertion logic similar to Separation Logic~\cite{reynolds2002} for quantum computing so as to be able to reason about only the interesting portions of the quantum state while still ensuring correctness of non-local effects such as entanglement. Various Quantum Hoare Logics that exist~\cite{ying2012} currently do not support frame rules that provide Separation Logic its power.\footnote{Since the time of this writing, we learned about Unruh's work~\cite{unruh2019} that we use in subsequent work~\cite{singhal2020quantum}.}

\paragraph{Linearity} Peter Selinger and collaborators have recently proposed a linear dependent typed version of Proto-Quipper (dubbed Proto-Quipper-D)~\cite{selinger2020,fu2020linear}. It is an interesting challenge to reconcile linearity in our theory based on their proposal.

\paragraph{Circuits as Arrows} Further, Proto-Quipper treats quantum circuits as first class citizens of the language. We would like to explore modifying our theory to treat Quantum Hoare types as arrows instead of as monads as was suggested by Vizzotto et al~\cite{so-arrows}. It makes sense from the perspective of sequential composition as arrows can have an arbitrary number of input/outputs as opposed to monads.

\paragraph{Behavioral Types} Another venue for exploration is to incorporate more precise types that can distinguish between qubits in pure classical state vs. those in superposition vs. those in entanglement~\cite{JorrandPerdrix2009} such as those inspired by the various quantum resource theories or the Heisenberg representation of quantum mechanics~\cite{rand2019,rssl20}. This may help us provide more specific postconditions that quantum programmers expect to hold true of their programs.

\paragraph{Classical Effects} Finally, it will be an interesting challenge to reconcile both classical and quantum effects together in a single grand unified theory for effects.

\subsection*{Acknowledgments}
We thank Robert Rand and the anonymous reviewers for their feedback on a previous draft of this paper.
This work is funded by EPiQC, an NSF Expedition in Computing, under grant CCF-1730449.

\bibliographystyle{eptcsalphaini}
\bibliography{ref}

\appendix

\section{Modular Bell Pair}
\label{app:mbp}
Writing a modular version of \texttt{testBell}.

\subsection{Hadamard basis states}
\begin{lstlisting}[language=QHaskell]
qplus : {emp} r : Qbit {Id(r, |+\>)}
      = do q <= mkQbit false;
           applyU (H q);
           return q

qminus : {emp} r : Qbit {Id(r, |-\>)}
       = do q <= mkQbit true;
            applyU (H q);
            return q
\end{lstlisting}

\subsection{Creating entanglement}
\begin{lstlisting}[language=QHaskell]
share : \Pi a : Qbit.
        {a \in {|+\>, |-\>}}   -- a \in {b, c} is short for Id(a,b) \/ Id(a,c)
           b : Qbit
        {Id(a, b) /\ a \in {|0\>, |1\>}}
      = \a.do b <= mkQbit false;
              applyU (ifQ a (X b));
              return b
\end{lstlisting}

\subsection{Bell pair}
\begin{lstlisting}[language=QHaskell]
bell : {emp} (a, b) : (Qbit, Qbit) {Id(a, b) /\ a \in {|0\>, |1\>}}
     = do qa <- qplus;
          qb <- share qa;
          return (qa, qb)
\end{lstlisting}

\subsection{Testing modular Bell pair}
\begin{lstlisting}[language=QHaskell]
testBell : {emp} (a, b) : (Bool, Bool) {emp /\ Id(a, b)}
         = do (qa, qb) <- bell;
              (measQbit qa, measQbit qb)
\end{lstlisting}

\section{Teleportation}
\label{app:teleport}
\subsection{Alice's circuit}
\begin{lstlisting}[language=QHaskell]
alice : \Pi a : Qbit. \Pi e : Qbit
        {(Id(a, -) /\ entangled(e)}
           r : (Bool, Bool)
        {emp}
      = \a.\e.do applyU (ifQ a (X e));
                 applyU (H a);
                 (measQbit a, measQbit e)
\end{lstlisting}

\subsection{Bob's circuit}
\begin{lstlisting}[language=QHaskell]
bob : \Pi m1 m2 : Bool. \Pi e : Qbit
      {entangled(e)}
         r : Qbit
      {(Id(r, -)}
    = \m1.\m2.\e.do if m1 then applyU (Z e) else ();
                    if m2 then applyU (X e) else ();
                    return e
\end{lstlisting}

\subsection{Teleport}
\begin{lstlisting}[language=QHaskell]
teleport : \Pi q : Qbit. x : Pure.
           {Id(q, x)}
              r : Qbit
           {Id(r, x)}
         = \q.do (a, b) <- bell;
                 (m1, m2) <- alice q a;
                 tq <- bob m1 m2 b;
                 return tq
\end{lstlisting}

\end{document}